\documentclass[twoside,fleqn]{article}
\usepackage{graphicx}
\usepackage{amsmath}
\usepackage{espcrc2}

\newcommand{\beq}{\begin{eqnarray}}
\newcommand{\eeq}{\end{eqnarray}}

\newcommand{\pardis}{\langle \mu \rangle}

\newcommand{\ie}{{\it i.e.\ }}

\title{Dual superconducting properties of the QCD vacuum}

\author{A. D'Alessandro\address[GENO]{Dipartimento di Fisica
 dell'Universit{\`a} di Genova and INFN, Via Dodecaneso 33, I-16146,
 Genova, Italy} and 
M. D'Elia\addressmark[GENO]\thanks{Speaker at the Conference}}

\begin{document}

\begin{abstract}
A consistent description of the confining QCD vacuum 
as a dual superconductor requires a determination of fundamental 
parameters such as the
superconductor correlation length $\xi$ and the field penetration
depth $\lambda$, which determine whether the superconductor is of
type I or type II.
We illustrate preliminary results of a lattice determination of $\xi$ 
for the case of pure Yang-Mills with two colors, obtained by measuring
the temporal correlator of a disorder parameter detecting dual superconductivity. 
\end{abstract}

\maketitle

\section{Introduction}
Color confinement is an absolute property of strongly interacting matter
which is still not explained from QCD first principles. 
Models exist, which relate confinement to some
property of the QCD fundamental state. 
One of those models, proposed by 't Hooft and Mandelstam,
is based on dual superconductivity of the vacuum~\cite{thooft75,mandelstam}:
confinement is related to the spontaneous breaking of a magnetic
symmetry produced by the condensation of some magnetically charged 
Higgs field. 
The magnetic condensate filling the QCD vacuum repels electric
fields out of the medium (dual Meissner effect), thus leading 
to the formation of flux tubes between colored charges, to the
linearly rising potential, and to confinement.
The broken magnetic group is chosen by a procedure known as
{\it Abelian projection}~\cite{thooft81}: 
a local operator $\phi (x)$ transforming
in the adjoint representation is diagonalized, leaving  
a residual $U(1)^{N_c-1}$ gauge symmetry.

The description of the QCD vacuum 
as a dual superconductor requires a determination of some fundamental 
parameters which characterize  it.
These are the correlation length $\xi$ of the Higgs condensate, and the field 
penetration depth $\lambda$. They determine whether the superconductor
is of type I ($\xi > \lambda$) or type II ($\xi < \lambda$). 
In a superconductor of type I an 
external field $B$ is always expelled from the medium  till a critical
value $B_c$ beyond which superconductivity disappears. In a superconductor
of type II there are instead two different critical values $B_{c1}$
and $B_{c2}$, and for $B_{c1} < B < B_{c2}$ the external field 
can penetrate the medium in the form of Abrikosov flux tubes, without
disrupting superconductivity. In a type II superconductor there is
repulsive interaction between two parallel flux tubes, which is
instead  attractive for type~I.

Whether the QCD vacuum behaves as a type I or type II dual
superconductor is a question which can be answered by numerical
lattice simulations. 
A direct way to determine $\lambda$ 
is a lattice analysis of the flux tube which is formed between two 
static color charges: this is done by measuring the chromoelectric
field correlated to a temporal Wilson loop. The longitudinal field
is expected to decay exponentially at large distances from the 
flux tube axis, as dictated by the field penetration depth
$\lambda$. Several studies have been
done~\cite{suz94,cea95,bali98,suz99,suz03,hay05},
giving $\lambda \sim 0.16$ fm for the $SU(2)$ pure gauge theory.

Determinations of $\xi$ instead have usually been based 
on an indirect analysis of the chromo-electric field distribution 
in the flux tube: 
$\xi$ is determined either through an analysis of violations
of the London equation close to the center of the flux tube~\cite{cea95,hay05}
or through some global fit to the whole set of Ginzburg-Landau
equations~\cite{bali98,suz99,suz03}.
Some assumption is needed anyway and the determinations
are not always consistent. An approximate picture emerges placing the 
$SU(2)$ QCD vacuum roughly at the boundary between 
a type I and type II dual superconductor.
A determination of $\xi$ through some observable directly related to 
the Higgs field would probably clarify
the issue: this has been the subject of our study.
The observable chosen in our work is the operator $\mu$ developed by the 
Pisa group which creates a magnetic monopole~\cite{dig97}.
Its {\em v.e.v.} $\langle \mu \rangle$ is a good disorder parameter detecting 
dual superconductivity ($\langle \mu \rangle \neq 0$) and
the transition to the deconfined - normal conducting phase
($\langle \mu \rangle = 0$) both in pure gauge 
theory~\cite{artsu2,artsu3,artran} and in full QCD~\cite{artfull,artfull2}.
Another approach, also aimed at a direct determination of $\xi$ but
by using a different observable, 
has been followed in Ref.~\cite{suz05}, which has 
appeared at the stage of writing these proceedings up.

\section{The method}

The operator $\mu$ is defined in the continuum as 
\begin{equation}
\mu^a (\vec x, t) =  
 e^{\,\,\, i \int\!\! d\vec y \,\,{\rm Tr} \{ \phi^a(\vec y, t) \vec
 E (\vec y, t)\} \vec b_\perp (\vec x - \vec y) } \label{MUA} \,
\end{equation}
where $\phi^a(\vec y, t)$ is the adjoint field defining the abelian
projection, $ \vec b_\perp $ is the field of the monopole
sitting at $\vec x$.
On the lattice correlation functions of $\mu (\vec x, t)$ are written as 
\beq
\langle \bar{\mu}(t',\vec{x}')\mu(t,\vec{x}) \rangle = {\tilde Z}/{Z}
\eeq
where $Z$ is the usual QCD partition function and $\tilde Z$ is 
obtained by changing the usual pure gauge action 
$S \to \tilde S$ by insertion of the monopole field in temporal
plaquettes at slices $t$ and $t'$.
Instead of $\langle \bar\mu \mu \rangle$ the following quantity is
measured
\beq
\rho = \frac{d}{d \beta} \ln \langle \bar\mu \mu \rangle = 
 \langle S \rangle_S - 
\langle \tilde{S} \rangle_{\tilde{S}} \, .
\eeq

The correlation length of $\mu$, $\xi_\mu$, can be obtained
by measuring the mass $M$ of the lowest state coupled to $\mu$, 
$\xi_\mu = M^{-1}$. This can be done by studying the 
asymptotic behaviour of the temporal correlator~\cite{dig97}
\begin{eqnarray}
\langle \bar{\mu}(t,\vec{x})\mu(0,\vec{x}) \rangle \simeq
  \langle \mu \rangle^2 + \gamma e^{-M t} \, .
\end{eqnarray}

In the confined phase $\pardis \neq 0$ , so for large $t$ the 
exponential decaying term is small
and we obtain (the hat denotes adimensional lattice quantities)
\begin{eqnarray}
\rho(\hat{t}) &\simeq& \frac{d}{d \beta} \left( \ln(\langle \mu \rangle^2) 
 + \frac{\gamma}{\langle \mu \rangle^2} e^{-\hat{M} \hat{t}} \right) = 
\nonumber \\ 
&=& A + (B + C \hat{t}) e^{-\hat{M} \hat{t}} 
\end{eqnarray}
$$A =  \frac{d \ln(\langle \mu \rangle^2)}{d \beta}\,; 
\,\,\,\,\, B = \frac{d}{d \beta} \frac{\gamma}{\langle \mu \rangle^2}\,;
\,\,\,\,\, 
C = - \frac{\gamma}{\langle \mu \rangle^2} \frac{d \hat{M}}{d \beta} $$
Since
\beq
\rho(\hat{t}) = \langle S \rangle_S - 
\langle \tilde{S}(\hat t) \rangle_{\tilde{S}(\hat t)}
\eeq
and $\langle S \rangle_S$ is independent of $\hat{t}$, we get
\beq
\langle \tilde{S}(\hat t) \rangle_{\tilde{S}(\hat t)} = 
A' + (B + C \hat{t}) e^{-\hat{t} a / \xi_\mu} 
\label{eqfit}
\eeq
We can thus obtain $\xi_\mu \equiv M^{-1}$ through the measurement of the 
expectation value 
$\langle \tilde{S}(\hat t) \rangle_{\tilde{S}(\hat t)}$
as a function of $\hat t$.

Before showing our numerical results, it is ne\-ces\-sary to discuss some
relevant questions. How is related the so measured $\xi_\mu$ 
to the actual correlation length of the Higgs field, $\xi$?
We do not really know which is the Higgs field which condenses in the
QCD vacuum, but we know that 
$\langle \mu \rangle$ is a good disorder parameter detecting dual 
superconductivity, therefore it is surely coupled to the Higgs field. 
$M$ is the mass of the lowest state coupling to $\mu$, we have
therefore $\xi_\mu = M^{-1} \geq \xi$.
Our investigation can then give a definite answer on the type
of dual superconductivity at work in the QCD vacuum only if 
we get $\xi_\mu < \lambda$, which implies $\xi < \lambda$,
\ie superconductivity of type II.
The result $\xi_\mu > \lambda$ would not lead to a definite
conclusion.

The next issue to be discussed regards the possible dependence of 
$\xi_\mu$ on the abelian projection chosen to define $\mu$.
The natural physical expectation is that one only coherence length
characterize the QCD vacuum.
This is consistent with 't Hooft ansatz that all abelian projections
are equivalent to each other: that equivalence also emerges
clearly from numerical determinations of 
$\langle \mu \rangle$~\cite{artsu2,artsu3,artran}.
A possible theoretical argument is the following: the operator
$\mu$ defined in one particular abelian projection creates
magnetic charge in \hbox{every} other abelian projection~\cite{abelind,artfull2};
this implies
that the lowest mass state coupled to $\mu$ should be universal,
\ie $\xi_\mu$ should be indepedent of the abelian projection chosen.
We will use a definition of $\mu$ which averages over different
abelian projections~\cite{artran}.

\section{Numerical Results}

\begin{figure}[hbt!]
\includegraphics*[angle=-90,width=\columnwidth]{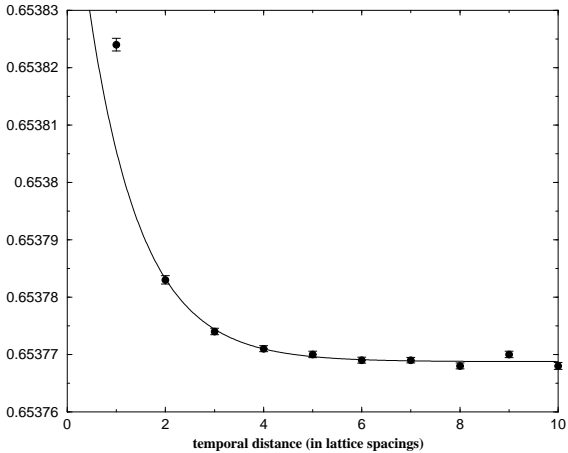}
\caption{$\langle \tilde{S}(\hat t) \rangle_{\tilde{S}(\hat t)}/6V$ as a
 function
of $\hat{t}$ at $\beta = 2.5115$ on a $16^3 \times 20$ lattice. The
 continuous
line is the result of a fit to Eq.~(\ref{eqfit}) where the first point
at $\hat{t} = 1$ has been discarded.}
\label{fig1}
\end{figure}

We first show results obtained for $SU(2)$ pure gauge theory
using the standard plaquette action at $\beta = 2.5115$:
this is the case which has been mostly studied in previous
literature.
The determination is made quite  difficult by the fact that 
the signal/background
ratio is very small: a precision of order $10^{-6}$ is necessary
in order to resolve the exponentially decaying term in Eq.~(\ref{eqfit}).
Some technical improvements, like link integration, have
been used in order to improve the precision on 
$\langle \tilde{S}(\hat t) \rangle_{\tilde{S}(\hat t)}$.
In Fig.~\ref{fig1}
we report the data for $\langle
\tilde{S}(\hat t) \rangle_{\tilde{S}(\hat t)}/6 V$ as a function of
$\hat t$ measured on a $16^3 \times 20$
lattice.

\begin{figure}[hbt!]
\includegraphics*[angle=-90,width=\columnwidth]{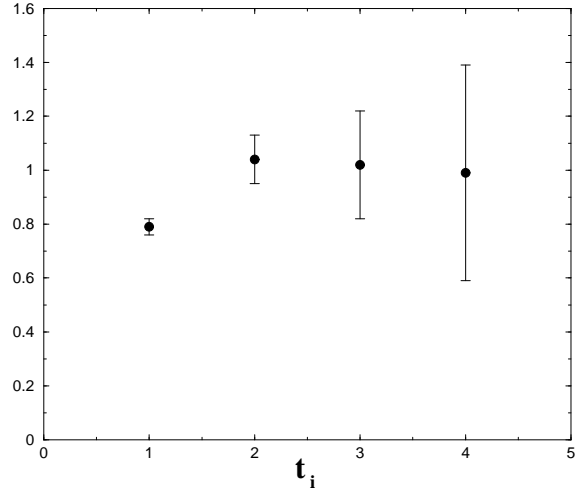}
\caption{Best fit results for $\xi_\mu/a$ at $\beta = 2.5115$ on 
a $16^3\times 20$ lattice as the starting point
$\hat t_{i}$ is changed. Results are stable for $\hat t_i \geq 2$. }
\label{fig2}
\end{figure}

A fit to the form of Eq.~(\ref{eqfit}) gives $C \simeq 0$,
\ie the linear term is not visible.
In Fig.~\ref{fig2} we show the best fit result for $\xi_\mu/a$ as the 
starting point $\hat t_{i}$ is varied: the stability of the results
is the signal that the asymptotic large $\hat t$ behaviour has
been reached.
A good and stable fit is obtained for
$\hat t_{i} \geq 2$ ($\chi^2/{\rm d.o.f.} = 1.3$ at $\hat t_{i} = 2$). 
From our fits we infer $\xi_\mu/a = 1.03 \pm 0.10$. Using 
$a(\beta = 2.5115) \simeq 0.091$ fm, as inferred from the 
non-perturbative determination of the $\beta$-function for $SU(2)$ 
reported  in Ref.~\cite{scalesu2}, we obtain 
$\xi_\mu = 0.094 \pm 0.009$ fm. As discussed above $\xi \leq
\xi_\mu$, therefore this result, when compared with the value 
$\lambda \sim 0.16$ fm reported in previous literature, gives indication
for dual superconductivity of type II. Our result for $\xi$ is
consistent
with those reported in Ref.~\cite{suz05}.

The creation operator 
$\mu$ brings in the magnetic field of a monopole. This is a long range
field and finite lattice size effects
could then be important.
We have therefore repeated our measurement on a $12^3\times20$
lattice, obtaining $\xi_\mu/a = 1.09 \pm 0.13$:
no evident difference is observed, thus excluding 
relevant finite size effects.

In order to check the scaling to the continuum of our results, 
we have also repeated our determination for a diferent value of
$\beta$, $\beta = 2.4$.
We have used a lattice size $12^3 \times 16$, which taking into
account $a(\beta = 2.4) \simeq 0.132$ fm~\cite{scalesu2} is comparable in
physical units to the largest one used at $\beta = 2.5115$. 
We have obtained $\xi_\mu = 0.11 \pm 0.01$ fm, which is 
compatible with the value at $\beta = 2.5115$.

\section{Conclusions}
We have investigated the nature of the dual superconducting vacuum
in $SU(2)$ pure gauge theory by determining the superconductor
correlation length $\xi$ through a measurement of the temporal correlator 
$<\bar\mu(t,\vec x) \mu(0,\vec x)>$ of a disorder parameter detecting
dual superconductivity.
 
Our determination, when compared with previous 
determinations of the 
dual field penetration depth $\lambda$, give indication
of weak type II dual superconductivity for the QCD vacuum.   

This result must be confirmed both by an increment of statistics
and by a wider study of the scaling to the continuum; a direct 
test of the abelian projection independence should be also performed.

\section*{Acknowledgements}
We acknowledge useful discussions with A.~Di~Giacomo. 
Numerical simulations have been performed on a PC cluster at 
INFN in Genova.

\end{document}